\documentclass[conference]{IEEEtran}
\IEEEoverridecommandlockouts
\usepackage{cite}
\usepackage{amsmath,amssymb,amsfonts}
\usepackage{algorithmic}
\usepackage{graphicx}
\usepackage{textcomp}
\usepackage{xcolor}
\usepackage{marvosym}

\usepackage{bbding}
\usepackage{booktabs}
\usepackage{multirow}
\usepackage{color}
\usepackage{hyperref}
\hypersetup{
    colorlinks=true,
    linkcolor=blue,
    filecolor=blue,      
    urlcolor=blue,
    citecolor=cyan,
}

\def\BibTeX{{\rm B\kern-.05em{\sc i\kern-.025em b}\kern-.08em
    T\kern-.1667em\lower.7ex\hbox{E}\kern-.125emX}}
\begin{document}

\title{
Collaborative Word-based Pre-trained Item Representation for Transferable Recommendation
\thanks{\Letter\ Corresponding author.}
\thanks{This work is supported by the Natural Science Foundation of China (Grant No. U21B2026, 62002191) and Quan Cheng Laboratory (Grant No. QCLZD202301).}
}
\author{\IEEEauthorblockN{Shenghao Yang\footnote{author}\textsuperscript{1,2,3,4}, Chenyang Wang\textsuperscript{1,3,4}, Yankai Liu\textsuperscript{4,6}, Kangping Xu\textsuperscript{1}, Weizhi Ma\textsuperscript{5}, Yiqun Liu\textsuperscript{1,3},\\ Min Zhang\textsuperscript{1,3,4,\Letter}, Haitao Zeng\textsuperscript{4,6}, Junlan Feng\textsuperscript{6} and Chao Deng\textsuperscript{6}}
\IEEEauthorblockA{
\textit{\textsuperscript{1}Department of Computer Science and Technology, Tsinghua University}\\
\textit{\textsuperscript{2}Quan Cheng Laboratory}
\textit{\textsuperscript{3}Zhongguancun Laboratory}\\
\textit{\textsuperscript{4}THU-CMCC Joint Institute}
\textit{\textsuperscript{5}AIR, Tsinghua University}
\textit{\textsuperscript{6}China Mobile Research Institute}\\
\{ysh21,xukp20\}mails.tsinghua.edu.cn, thuwangcy@gmail.com, \\ \{liuyankai,zenghaitao,fengjunlan,dengchao\}@chinamobile.com, \{mawz,yiqunliu,z-m\}@tsinghua.edu.cn
}
}

\maketitle

\begin{abstract}
Item representation learning (IRL) plays an essential role in recommender systems, especially for sequential recommendation.
Traditional sequential recommendation models usually utilize ID embeddings to represent items, which are not shared across different domains and lack the \textit{transferable} ability.
Recent studies use pre-trained language models (PLM) for item text embeddings (text-based IRL) that are universally applicable across domains.
However, the existing text-based IRL is unaware of the important collaborative filtering (CF) information.
In this paper, we propose CoWPiRec, an approach of \underline{Co}llaborative \underline{W}ord-based \underline{P}re-trained \underline{i}tem representation for \underline{Rec}ommendation.
To effectively incorporate CF information into text-based IRL, we convert the item-level interaction data to a word graph containing word-level collaborations.
Subsequently, we design a novel pre-training task to align the word-level semantic- and CF-related item representation. 
Extensive experimental results on multiple public datasets demonstrate that compared to state-of-the-art transferable sequential recommenders, CoWPiRec achieves significantly better performances in both fine-tuning and zero-shot settings for cross-scenario recommendation and effectively alleviates the cold-start issue.
The code is available at: \url{https://github.com/ysh-1998/CoWPiRec}.
\end{abstract}

\begin{IEEEkeywords}
Recommender System, Item Representation Learning, Transfer Learning  
\end{IEEEkeywords}

\section{Introduction}
Item representation learning (IRL) is a crucial technology in recommender systems since items interacted by users largely reflect their preferences.
IRL is especially important for sequential recommendation, where user representations are typically obtained by aggregating the representations of interacted items~\cite{hidasi2015session,wang2019sequential}. 
Specifically, sequential recommender comprises two main components: the IRL module used to obtain item representations, and the sequence representation learning (SRL) module used to aggregate the representations of the chronologically-ordered items. 
Recent neural sequential recommendation models typically use an ID-based IRL module to map item IDs to hidden vectors and an SRL module with advanced neural networks, e.g., transformer layers~\cite{kang2018self}.
Then the two modules are trained simultaneously with optimization objective of the next-item prediction task~\cite{hidasi2015session,kang2018self}. 
Although promising results have been achieved, these methods heavily rely on rich ID-based interactions. 
When new scenarios arise, the models need to be trained from scratch since the ID embeddings are not shared across scenarios and may suffer the cold-start issue. 
Therefore, sequential recommendation models with ID-based IRL lack the \textit{transferable} ability.

Recently, many content-based sequential recommendation models have been proposed to alleviate the above issue. 
Especially, considering the generalization of the text and the cross-scenario shared vocabulary, many works use the representation of item text instead of the ID embedding, i.e., text-based IRL. 
Due to the remarkable performance of pre-trained language model (PLM)~\cite{devlin2018bert} in neural language processing, existing works typically use PLM as the text-based IRL module. 
Specifically, these works obtain text-based item representations offline with PLM and feed the item representations into the SRL module. 
Then the SRL module is pre-trained on mixed-domain data to learn cross-domain general sequential representation patterns and the learned knowledge is transferred to a new domain, resulting in transferable sequential recommender~\cite{ding2021zero,hou2022towards}. 


However, Although text-based item representations have effective semantic representation capabilities, they do not contain collaborative filtering (CF) information. 
In fact, some words that are not similar in semantics might be closely related in the context of recommendation. 
For example, ``health” and ``cycling” are two words that are not very close in terms of semantic representation space. 
While in the recommendation scenario, a user interested in healthy food may also prefer to buy some cycling equipment for exercise. 
To alleviate this issue, we argue that it is desired to incorporate CF-related signals into the text-based IRL.
While most existing approaches focus on pre-training the SRL module and the PLM is frozen in training and unaware of important CF signals. 


In this paper, we propose a \textbf{Co}llaborative \textbf{W}ord-based \textbf{P}re-trained \textbf{i}tem representation for \textbf{Rec}ommendation, \textbf{CoWPiRec}.
Specifically, we extract word-level CF signals, i.e. co-click words, from user interaction history and construct a word graph to integrate these co-click relationships. 
Subsequently, we design a novel word-level pre-training task to incorporate CF signals into PLM. The word graph serves as a CF-related knowledge source to instruct the pre-training procedure.

The merits of our proposed item representation learning approach are threefold. 
Firstly, since CoWPiRec is pre-trained independent of the SRL module, it is convenient to be integrated into different sequence aggregation networks as the text-based IRL module. 
Secondly, the item representation generated by CoWPiRec provides both effective semantic matching and CF-related signals, it can be used to perform recommendation tasks without any training stage when transferring to a new domain, i.e.,  zero-shot recommendation.
Thirdly, CoWPiRec further achieves outperforming recommendation results with in-domain training utilizing the CF-related knowledge learned in pre-training.

We evaluate the effectiveness of CoWPiRec in the cross-scenario setting. 
We first use datasets from multiple domains to construct the word graph and pre-train CoWPiRec. 
Then, considering the efficiency in a new scenario, we utilize CoWPiRec as a feature extractor to offline generate item representations.
The item representations can be used to perform downstream recommendations.
The experiment results on the public datasets demonstrate that CoWPiRec outperforms state-of-the-art approaches in the zero-shot recommendation and further improves in-domain training effectiveness.
%

The main contributions of our work are summarized as follows:
\begin{itemize}
  \item We propose a pre-trained item representation learning approach that aligns semantic and collaborative information for the recommendation.
  \item We design a novel pre-training task to incorporate word-level CF signals from the co-click word graph into the text-based IRL. 
  \item Comparative experimental results on multiple public datasets demonstrate that our proposed approach achieves significantly better performances and effectively alleviates the cold-start issue.
\end{itemize}

\section{Related Work}
\subsection{Sequential Recommendation}
Sequential recommendation is a widely researched topic in the recommendation system community, with the objective of predicting the next item of a user's interaction history~\cite{wang2019sequential,kang2018self}. 
Early studies are based on Markov chain assumptions to estimate the transition relationships between items~\cite{rendle2010factorizing,hidasi2016general}. 
In recent years, with the development of deep learning, neural sequential recommendation models based on deep neural networks have emerged. 
These models usually comprise item representation learning (IRL) and sequence representation learning (SRL) modules to model the representation of item and user sequences.
The SRL module utilizes various network structures, including Recurrent Neural Networks (RNN)~\cite{hidasi2015session,li2017neural,jang2020cities}, Convolutional Neural Networks (CNN)~\cite{tang2018personalized}, Transformer~\cite{kang2018self,10027680,he2021locker,sun2019bert4rec,hou2022core}, and Graph Neural Networks (GNN)~\cite{chang2021sequential,wu2019session,xiao2023social4rec}, to modeling the user sequence representation by aggregating the item representations.
The item representations are obtained with the IRL module.
Most IRL modules utilize item ID embedding to map item ID to a hidden vector~\cite{hidasi2015session,kang2018self}.
Limited by unshareable item IDs, these approaches with the ID-based IRL module lack transferable ability across scenarios.
Different from relying on explicit item IDs, we represent items based on item text to enhance the transferable ability of sequential recommender.

\subsection{Recommendation with Pre-trained Language Model}
Inspired by the rapid development of the pre-trained language model (PLM), many recent works use PLM as the IRL module of the recommendation model ~\cite{zhang2021unbert,wu2021empowering,ding2021zero,yu2022tiny,hou2022towards,hou2022learning,wang2022transrec,mu2023id,yuan2023go}. 
With semantically enhanced item representations, these approaches achieve significant performance improvement in the recommendation and effectively alleviate the cold-start issue. 
These works can be divided into two main lines. 
One line is to perform joint training of PLM and the SRL module to adapt to the recommendation tasks.
PLM-NR~\cite{wu2021empowering} utilizes PLM and an attention network to obtain item text representations. Then perform joint training on the SRL module and the last two layers of the PLM in the news recommendation.
Due to the high computation complexity of PLM, another line is to generate item text representations offline with PLM.
IDA-SR~\cite{mu2023id} utilizes PLM to obtain the item representations as input to the SRL module. 
Subsequently, three pre-training tasks are used to bridge the gap between text semantics and sequential user behaviors.
Works of this line only train the SRL module and PLM is unaware of task-specific information, which leads to a suboptimal performance.
Considering performance and efficiency tradeoffs, our approach trains PLMs in the pre-train stage to learn CF-related knowledge.
When transferring to a new domain, we use the tuned PLMs to generate item representations offline, thus improving efficiency.

\subsection{Transferable Recommendation Systems}
Improving the transferable ability of recommender systems is a rapidly growing research area.
It aimed at leveraging knowledge learned from multiple domains to enhance the performance of the recommendation model in new domains~\cite{zhu2021cross,xu2022centralized}. 
Early studies typically assume the presence of commonalities across various domains, such as users with similar preferences~\cite{hu2018conet,wu2020ptum,xiao2021uprec,yuan2021one} and common items\cite{singh2008relational,zhu2019dtcdr}, to enable mapping between the source and target domains.
Recent works have attempted to achieve transferable sequential recommender by learning cross-domain universal representations\cite{ding2021zero,hou2022towards,shin2021scaling,wang2022transrec,hou2022learning}.
ZESRec~\cite{ding2021zero} utilizes the universal item text representations obtained by PLM and performs the next item prediction task on the SRL module.
The trained SRL module could transfer to a new domain with the item text representations as input.
UniSRec~\cite{hou2022towards} further adapt item text representations with an MoE module and enables the SRL module to learn a universal sequence pattern with the sequence-item and sequence-sequence contrastive pre-training tasks.

Most existing works focus on pre-training a transferable SRL module and the PLM is frozen. 
The item representation obtained by PLM can only provide semantics information and lacks CF-related signals, which limits the overall performance.
To address this issue, we propose to incorporate recommendation signals into PLM via CF-related tasks.
MoRec~\cite{yuan2023go} is a recently proposed work with an idea close to ours. It performs a joint training of the PLM and the SRL module with a next-item-prediction task.
However, since PLM is typically pre-trained with the word-level task, e.g., masked language modeling~\cite{devlin2018bert}, the supervision signals of item-level recommendation tasks don't match PLM well.
To align with the modeling strategy of PLM, we incorporate word-level CF signals into PLM through a word-level pre-training task.

\begin{figure*}[t]
  \centering
  \includegraphics[width=0.95\linewidth]{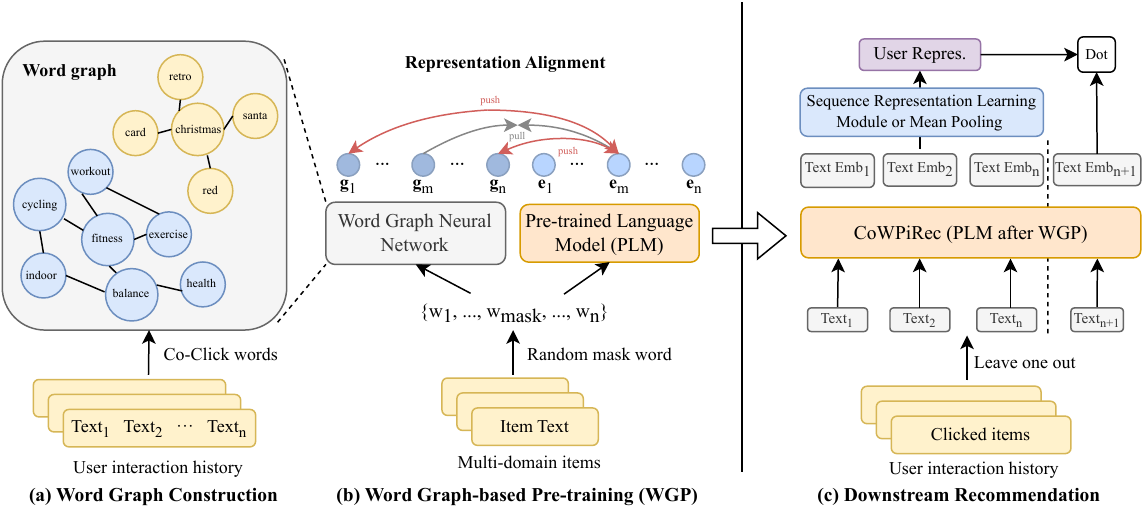}
  \caption{The overall framework of our proposed collaborative word-based pre-trained item representation for recommendation (CoWPiRec). (a) The word-level collaborative filtering (CF) signals are from the co-click relationships of word pairs in the word graph. (b) A word graph-based pre-training (WGP) is performed to align the semantic- and the CF-related representation of the PLM and word graph with contrastive learning. $w_i$ denotes the word in item text, $\mathbf{g}_i$ and $\mathbf{e}_i$ is the word representation after word graph modeling and semantic modeling (c) When transferring CoWPiRec to a new domain, the item representations generated offline utilizing CoWPiRec are fed into a sequence representation learning (SRL) module or a simple mean-pooling to perform downstream recommendations in the fine-tuning and zero-shot settings, respectively.}
  \label{fig_model}
\end{figure*}
\section{Methodology}
In this section, we present our proposed transferable item representation learning approach, CoWPiRec.
Utilizing the word-level CF knowledge learned from the co-click word graph, CoWPiRec generates item representation with both semantic and CF-related information based on item text.
When transferring to a new domain, the enhanced item representation could directly perform recommendations without training procedure and contribute to the in-domain training.
\subsection{Framework Overview}
The overall framework of our proposed text-based IRL approach is shown in Figure~\ref{fig_model}.
Text-based IRL approach utilizes item text representation generated by PLM to replace the ID-based item representation of traditional sequential recommendation models. 
It has achieved promising transferable recommendation performance combined with a pre-training scheme on the SRL module~\cite{ding2021zero,hou2022towards}. 
We argue that these transferable recommenders are suboptimal since the text-based IRL modules are unaware of CF-related information and it is desired to incorporate CF-related signals into PLM.

Considering PLM is typically trained with the word-level task, the item-level next-item-prediction task is not applicable to integrate CF signals into PLM.
Therefore, we first extract word pairs with co-click relationships from interaction data and construct a word graph that contains these relationships.
The co-click relationships between these words can be seen as word-level CF signals.
Then we incorporate the word-level CF signals from the word graph into PLM through a word-level pre-training task.
We will explain each key component of our proposed approach in the following sections.

\subsection{Word Graph Construction}
In this section, we present the process of extracting co-click words and constructing the word graph. A sub-graph of our constructed word graph is shown in Figure~\ref{fig_model} (a). The co-click relationship is a common concept in recommender systems while previous works mostly focus on item-level co-click relationships.
To align with the modeling format of PLM and incorporate the recommendation signal more effectively, we extract the word-level co-click relationships from the item text.

In different recommendation scenarios, although items have different presentation formats, they usually have basic textual descriptions.
Due to the universality of language, different domains share a common vocabulary, making the text bridge different recommendation scenarios. 
Additionally, item texts often contain some word-level user preferences. 
If a user clicks several items containing words like ``health” or ``fitness”, it indicates that this user may be focused on a healthy lifestyle.
Therefore, the user may be also interested in nutritionally balanced food or some fitness equipment. 
These items may contain words such as ``balance”, ``exercise” and ``cycling”. 

We construct a word graph to organize the co-click relationships based on user interaction. 
For each word, a candidate set of words is generated based on co-click relationships and then filtered to retain only the top $N$ words as neighboring nodes.

Specifically, given a user's interaction sequence $s=\{i_1,i_2,i_t,...,i_n\}$, where $i_t$ represents the $t$-th item in the sequence. 
A co-click word pair is defined as two words from each item text respectively, denoted as $w_i$ and $w_j$. 
We count the occurrences of all co-click word pairs, denoted as $(w_i, w_j, c_{ij}, c_{ji})$, where $c_{ij}=c_{ji}$. 
Since each word contains a large number of co-click words, we follow~\cite{shi2021wg4rec} and filter the candidate co-click words using the $tf\mbox{-}idf$ algorithm. 
The $tf\mbox{-}idf$ value of a pair of co-click words is calculated by Equation~\eqref{tf-idf}:
\begin{equation}
\begin{gathered}
\label{tf-idf}
    tf_{i,j}=\frac{c_{i,j}}{\sum_{k=1}^{V} c_{i,k}}, \quad idf_{j}=\lg \frac{V}{\left|\left\{c_{k,j} \mid \forall k, c_{k,j}>0\right\}\right|}, \\
    tf\mbox{-}idf_{i,j} = tf_{i,j}\times idf_{j},
\end{gathered}
\end{equation}
where $V$ is the vocabulary size, and the denominator of $idf$ is the number of words that have the co-click relationship with $w_j$. 
The higher the $tf\mbox{-}idf$ value, the more times $w_j$ and $w_i$ are co-clicked and the less $w_j$ is co-clicked with other words. 
For each $w_i$, only the top $N$ words with the highest $tf\mbox{-}idf$ values will be selected as its neighbor nodes in the word graph.

By constructing edges between co-click word pairs, we obtain a word graph fusion of word-level CF signals.
We construct the word graph based on the user interaction data of multiple domains to improve the generality ability of extracted word-level CF signals. The word pairs with edges in the word graph may relate to different domains, e.g. ``health” and  ``balance” in the ``Food” domain, ``cycling”, ``indoor” and ``exercise” in the ``Home” domain, as shown in Figure~\ref{fig_model} (a).

\subsection{Word Graph-based Pre-training Task}
With the remarkable semantic representation ability of PLM, text-based IRL based on PLM provides an effective semantic matching ability.
While PLM cannot capture CF-related information and this limits the representation ability of text-based IRL. 
To incorporate the recommendation signal into PLM, an intuitive idea is to train PLM and SRL simultaneously via the next item prediction task, thus introducing task-specific information into PLM. 
However, since PLM's modeling method on large-scale corpora is word-level, the above item-level supervision signal cannot be well integrated into PLM. 

Considering many works have demonstrated that aligning with PLM's modeling format in downstream tasks can better inspire its learned knowledge~\cite{liu2023pre}, we propose a word-level pre-training task to incorporate the word-level CF information contained in the word graph into the PLM, as shown in Figure~\ref{fig_model} (b).
Specifically, we use item text as the input of the PLM and add special symbols [CLS] and [SEP] before and after the input in accordance with the input format of the PLM. 
We randomly mask words in the item text using the [MASK] special symbol. 
For an item text input $i=\{cls, w_1,...,w_m,...,w_n, sep\}$, where $w_m$ is the masked word, the initialize word embedding of each word is obtained with PLM's word embedding, i.e., $\{\mathbf{v}_{cls}, \mathbf{v}_{1},..., \mathbf{v}_{m},..., \mathbf{v}_{n}, \mathbf{v}_{sep}\}$, where $\mathbf{v}_i\in \mathbb{R}^d$ and $d$ is the dimension of word embedding. 
Then two different modeling procedures are performed for input word embedding, namely \textit{semantic modeling} and \textit{word graph modeling}.

\subsubsection{Semantic Modeling} 
In this modeling procedure, the word embedding of each word in item text $\mathbf{v}_i$ is firstly concatenated, i.e., $\mathbf{x}=[\mathbf{v}_{cls}; \mathbf{v}_{1};...; \mathbf{v}_{m};...; \mathbf{v}_{n}; \mathbf{v}_{sep}]\in\mathbb{R}^{n\times d}$, 
where $n$ is the input length and we omit the special token at the head and tail for convenience.
$[;] $ is the concatenation operation.  
Then $\mathbf{x}$ is fed into the $L$-layer Transformer encoder of the PLM. Each Transformer encoder consists of a multi-head self-attention layer and a position-wise feed-forward layer. 
A residual connection and layer normalization are performed in the above two parts. 
We set $\mathbf{x}^0\in\mathbb{R}^{n\times d}$ as the input, and the output after $l+1$-layer Transformer encoder is obtained by Equation~\eqref{trm}.
\begin{equation}
\begin{gathered}
\label{trm}
\mathbf{x}^{l+1} = Trm(\mathbf{x}^{l}) = LN(\mathbf{s}^{l}+FFN(\mathbf{s}^{l})), \\
\mathbf{s}^{l} = LN(\mathbf{x}^{l}+MHAttn(\mathbf{x}^{l})),
\end{gathered}
\end{equation}
where $Trm(\cdot)$ is the Transformer encoder layer, $LN(\cdot)$ is the layer normalization function, $FFN(\cdot)$ is the position-wise feed-forward layer, $MHAttn(\cdot)$ is the multi-head self-attention layer, $\mathbf{s}^{l}\in \mathbb{R}^{n\times d}$ is the output of the multi-head self-attention layer. The output of the last layer is $\mathbf{x}^{L}=[\mathbf{e}_{cls}; \mathbf{e}_{1};...; \mathbf{e}_{m};...; \mathbf{e}_{n}; \mathbf{e}_{sep}] \in \mathbb{R}^{n\times d}$. 

With the self-attention mechanism of the Transformer Encoder, $\mathbf{e}_i\in \mathbb{R}^d$ integrates the contextual information of other words in the item text, which demonstrates effective semantic representation ability in many tasks.
While in the recommendation system, semantic similarity and recommendation relevance are not related, so the PLM is expected to capture additional recommendation signals to improve the recommendation performance.

\subsubsection{Word Graph Modeling} 
In this part, the representation of each word in input $\mathbf{x}$ is obtained by aggregating the embedding of its neighboring nodes through a graph neural network (GNN). 
Specifically, we follow~\cite{shi2021wg4rec} and use the GraphSAGE algorithm~\cite{hamilton2017inductive} to learn a function for aggregating neighbor node representations. 
\begin{equation}
\label{graphsgae}
\mathbf{h}_i^t=\sigma\left(\mathbf{W}_g\left(\mathbf{h}_i^{t-1} \oplus \operatorname{AGG}\left(\left\{\mathbf{h}_j^{t-1}, \forall w_j \in \mathcal{N}_{w_i}^*\right\}\right)\right)\right),
\end{equation}
where $\mathbf{h}_i^t\in \mathbb{R}^d$ is the representation of central word $w_i$ in the $t$-th layer of GNN, which is aggregated with the representation of itself $\mathbf{h}_i^{t-1}$ and its neighbors $\mathbf{h}_j^{t-1}$ in the $t-1$ layer. 
The initial representation of each word is the initialized word embedding, i.e., $\mathbf{h}_i^0 = \mathbf{v}_i$. $\sigma$ is a non-linear activation function. $W_g\in\mathbb{R}^{d\times2d}$ is the weight of a linear layer. 
$\mathcal{N}_{w_i}^*$ is the sampled neighbors. $\oplus$ is a concatenate operator. $\operatorname{AGG}$ is an aggregating function based on the attention mechanism. It aggregates the representation of neighbors with Equation~\ref{graphatt}.
\begin{equation}
\label{graphatt}
\begin{gathered}
\mathbf{q}_g^t=\sigma\left(\sum_{w_j \in \mathcal{N}_{w_i}^*} Q_g \mathbf{h}_j^{t-1}\right), \quad \mathbf{k}_j^t=\sigma\left(K_g \mathbf{h}_j^{t-1}\right) \\
a_j^t=\frac{\exp \left(\mathbf{q}_g^{t T} \boldsymbol{k}_j^t\right)}{\sum_{w_k \in \mathcal{N}_{w_i}^*} \exp \left(\mathbf{q}_g^{t^T} \boldsymbol{k}_k^t\right)}, \quad \mathbf{h}_{\mathcal{N}_{w_i}^*}^t=\sum_{w_j \in \mathcal{N}_{w_i}^*} a_j^t \mathbf{h}_j^{t-1},
\end{gathered}
\end{equation}
where $Q_g, K_g \in \mathbb{R}^{d \times d}$ is the weight of the projection layer and $a^t_j$ is the attention weight of each neighbor. The output after $T$ layers of GNN is
\begin{equation}
\label{gnn}
\begin{gathered}
\mathbf{g}_i = \mathbf{h}_i^T = \sigma(W_g(\mathbf{h}_i^{T-1}\oplus\mathbf{h}^{T-1}_{\mathcal{N}_{w_i}^*})).
\end{gathered}
\end{equation}
The central word representation $\mathbf{g}_i\in \mathbb{R}^d$ aggregated with co-click words is fused with the word-level CF signal of the word graph.

\subsubsection{Representation Alignment} In order to incorporate the word-level CF signals extracted from the word graph into the representation space of PLM, we adopt a widely used contrastive learning method to align the semantic representation of PLM $\mathbf{e}_i\in \mathbb{R}^d$ with the CF-related representation of word graph $\mathbf{g}_i\in \mathbb{R}^d$. Specifically, for a masked word $w_m$ in the input, we obtain its representations of PLM and word graph, i.e., $\mathbf{e}_m\in \mathbb{R}^d$ and $\mathbf{g}_m\in \mathbb{R}^d$. We treat them as a positive pair and treat $\mathbf{g}_i$ of other words in the same input ($i\ne m$) as negatives. We aim to pull $\mathbf{e}_m$ and $\mathbf{g}_m$ closer and push $\mathbf{e}_m$ away from other $\mathbf{g}_i$ by minimizing the following contrastive learning loss:
\begin{equation}
\label{cl}
\mathcal{L}=-\frac{1}{M} \sum_{m=1}^M \log \frac{\exp \left(\mathbf{e}_m \cdot \mathbf{g}_m / \tau\right)}{\sum_{i=0}^n \exp \left(\mathbf{e}_m \cdot \mathbf{g}_i / \tau\right)}, i\ne m,
\end{equation}
where $M$ is the number of masked words of the input item text.

It is worth noting that, during the training process, there is a parameter sharing between the word embedding of the PLM and the node embedding of the word graph. As a result, the output of a word in the PLM gradually approaches its aggregated representation of neighbor nodes in the word graph. This process results in the PLM's output containing both semantic information and word-level CF information. We refer to this recommendation-orient trained PLM as CoWPiRec.

\subsection{Downstream Recommendation}
Through constructing word graphs and pre-training on multiple domains, we obtain a text-based IRL module, CoWPiRec, that captures word-level CF signals. When transferring to a new domain, we consider two settings to evaluate the effectiveness of CoWPiRec: \textit{fine-tuning setting} and \textit{zero-shot setting}. The downstream recommendation pipeline is shown in Figure~\ref{fig_model} (c).
\subsubsection{Fine-tuning Setting}
In this setting, we train a sequential recommendation model using all training data in the new domain. Following the standard pipeline, given a user's click sequence $s=\{i_1,i_2,...,i_n\}$, for each $i_t=\{w_1,w_2,...,w_n\}$, it is fed into CoWPiRec after adding special symbols [CLS] and [SEP]. The item representation is obtained by Equation~\eqref{item_rep}.
\begin{equation}
\label{item_rep}
\mathbf{i}_t = CoWPiRec([cls; w_1; w_2; ...; w_n; sep]),
\end{equation}
where $CoWPiRec(\cdot)$ takes the representation of the [cls] position as the item representations $\mathbf{i}_t\in\mathbb{R}^d$. Then we follow~\cite{hou2022towards} and used an MoE module consisting of
multiple whitening networks to adapt the item representations and reduce the dimension, resulting $\widetilde{\mathbf{i}}_t \in \mathbb{R}^{d_V}$.

We adopt a widely used transformer network to aggregate the item representations. Specifically,  we sum the item representations and the absolute position embedding $\mathbf{p}_t\in \mathbb{R}^{d_V}$ as the input.
\begin{equation}
\mathbf{f}^0_t = \widetilde{\mathbf{i}}_t + \mathbf{p}_t.
\end{equation}
Then $\mathbf{F}^{0}=[\mathbf{f}_1^0;...;\mathbf{f}_n^0] \in \mathbb{R}^{n \times d_V}$ is fed into $L$ transformer layers, the output after $l+1$ layers is:
\begin{equation}
\mathbf{F}^{l+1} = FFN(MHAttn(F^l)).
\end{equation}
We take the $t$-th position hidden state of the last layer, i.e., $\mathbf{f}^L_n\in \mathbb{R}^{d_V}$ as the user representation $\mathbf{u}\in \mathbb{R}^{d_V}$. 

Note that since CoWPiRec already has the ability to capture recommendation signals, we don't need to update the parameters of CoWPiRec during training. Therefore we offline obtain all item representations, which significantly improves efficiency. For user representation $\mathbf{u}$, we calculate the score of candidate next item $i_{t+1}$ using the dot product:
\begin{equation}
\label{score}
score_{(i_{t+1}|s)} = Softmax(\mathbf{u} \cdot \widetilde{\mathbf{i}}_{t+1}).
\end{equation}
We use the cross-entropy loss for the next item prediction task during training. In the inference stage, we rank the items based on the dot product score.

\subsubsection{Zero-shot Setting}
In contrast to the cold-start problem, the objective of zero-shot recommendation is to determine whether a model has basic recommendation capabilities without any in-domain training. It can not be achieved with traditional ID-based recommendation models. Since the item representations generated by CoWPiRec have a remarkable semantic matching ability and could capture recommendation signals. Therefore, we directly use the nearest neighbor search with the dot product to perform the recommendation. Specifically, given all item representations in a user sequence $\{\mathbf{i}_1,\mathbf{i}_2,...,\mathbf{i}_n\}$ obtained by CoWPiRec with Equation~\eqref{item_rep}. We use mean-pooling to aggregate the item representations to obtain the user representation $\mathbf{u}$.
\begin{equation}
\mathbf{u} = \frac{1}{n}\sum_{t=1}^n\mathbf{i}_t.
\end{equation}
Then the score of the candidate item $i_{t+1}$ is calculated with Equation~\eqref {score} and we directly predict the next item according to the scores.

\subsection{Discussion}

In this section, we present the differences between our proposed CoWPiRec compared with other sequential recommendation models. 
The comparison focuses on the two components of sequential recommendation models, i.e., the IRL and SRL modules, and the model's transferable ability. 
The comparison results are shown in Table~\ref{tab:dis}.

\textbf{ID-based IRL approaches} such as SASRec~\cite{kang2018self} and BERT4Rec~\cite{sun2019bert4rec} obtain item representations with explicit item IDs. SASRec utilizes transformer layers to aggregate item ID representations and BERT4Rec performs a mask item prediction task to pre-train the bidirectional transformer layer. Since item IDs are not shared across scenarios, these approaches need to be trained from scratch when applied to new scenarios and lack transferable ability.
CoWPiRec does not rely on the item ID to perform recommendations and adopt a text-based IRL module. With the shared vocabulary across scenarios, CoWPiRec achieves transferable recommendations.

\textbf{Text-based IRL approaches} such as S\textsuperscript{3}Rec~\cite{zhou2020s3} incorporate item text representation as an auxiliary feature and perform self-supervised tasks to integrate the representation of sequence, item, and feature.
Since S\textsuperscript{3}Rec also utilizes the item id embedding, the pre-train task can only be performed in-domain. 
Different from S\textsuperscript{3}Rec, ZESRec~\cite{ding2021zero} and UniSRec~\cite{hou2022towards} purely use item text representations and perform a cross-domain pre-training on the SRL module.
The pre-trained SRL module can learn general sequence modeling patterns and contribute to the cross-scenario recommendations.
Instead of focusing only on pre-training the SRL module, MoRec~\cite{yuan2023go} train the text-based IRL and SRL module jointly with the next-item-prediction task.
We don’t pre-train the SRL module in our proposed approach and perform a word graph-based per-training task to obtain a transferable text-based IRL module, i.e., CoWPiRec.
\begin{table}[t] 
\caption{Comparison of different sequential recommenders.}
\label{tab:dis}
\begin{tabular}{lccccc}
\toprule
\multirow{2}{*}{Methods} & \multicolumn{2}{c}{Used information}             & \multicolumn{2}{c}{Pre-training on}                          & \multirow{2}{*}{Transferable} \\
\cmidrule(lr){2-3} \cmidrule(lr){4-5}
                         & ID & Text & Item & Sequence &                               \\
\midrule
SASRec                   &    \textcolor[RGB]{0,130,130}{\CheckmarkBold}      &    \textcolor[RGB]{210,0,60}{\XSolidBrush}      &   \textcolor[RGB]{210,0,60}{\XSolidBrush}      &    \textcolor[RGB]{210,0,60}{\XSolidBrush}      &    \textcolor[RGB]{210,0,60}{\XSolidBrush}      \\
BERT4Rec                 &  \textcolor[RGB]{0,130,130}{\CheckmarkBold} &  \textcolor[RGB]{210,0,60}{\XSolidBrush}  &    \textcolor[RGB]{210,0,60}{\XSolidBrush} &    \textcolor[RGB]{0,130,130}{\CheckmarkBold}          &     \textcolor[RGB]{210,0,60}{\XSolidBrush}                \\
S3Rec                    & \textcolor[RGB]{0,130,130}{\CheckmarkBold} & \textcolor[RGB]{0,130,130}{\CheckmarkBold}     &    \textcolor[RGB]{210,0,60}{\XSolidBrush}  & \textcolor[RGB]{0,130,130}{\CheckmarkBold}    &     \textcolor[RGB]{210,0,60}{\XSolidBrush}           \\
ZESRec                   & \textcolor[RGB]{210,0,60}{\XSolidBrush} &  \textcolor[RGB]{0,130,130}{\CheckmarkBold}   &    \textcolor[RGB]{210,0,60}{\XSolidBrush}       &  \textcolor[RGB]{0,130,130}{\CheckmarkBold}        &   \textcolor[RGB]{0,130,130}{\CheckmarkBold}             \\
UniSRec                  & \textcolor[RGB]{210,0,60}{\XSolidBrush} &  \textcolor[RGB]{0,130,130}{\CheckmarkBold}  &   \textcolor[RGB]{210,0,60}{\XSolidBrush}     &    \textcolor[RGB]{0,130,130}{\CheckmarkBold}      &      \textcolor[RGB]{0,130,130}{\CheckmarkBold}        \\
MoRec                  & \textcolor[RGB]{210,0,60}{\XSolidBrush} &  \textcolor[RGB]{0,130,130}{\CheckmarkBold}  &   \textcolor[RGB]{0,130,130}{\CheckmarkBold}     &    \textcolor[RGB]{0,130,130}{\CheckmarkBold}      &      \textcolor[RGB]{0,130,130}{\CheckmarkBold}        \\
\midrule
CoWPiRec                   & \textcolor[RGB]{210,0,60}{\XSolidBrush} &  \textcolor[RGB]{0,130,130}{\CheckmarkBold}  &  \textcolor[RGB]{0,130,130}{\CheckmarkBold}        &  \textcolor[RGB]{210,0,60}{\XSolidBrush}          &  \textcolor[RGB]{0,130,130}{\CheckmarkBold}    \\
\bottomrule
\end{tabular}
\end{table}
\section{Experiments}
In this section, we first introduce how to evaluate the transferable ability of CoWPiRec in cross-scenario settings and then present experimental results and analysis.

\subsection{Experiment Setup}
\subsubsection{Datasets}
We use mixed-domain user interaction data to pre-train CoWPiRec, and then use multiple downstream datasets to evaluate the transferable performance of CoWPiRec. The statistics of the dataset used are shown in Table~\ref{tab:data}.
\begin{itemize}
  \item \textbf{Pre-trained datasets}: We select the datasets from five domains in the Amazon dataset~\cite{ni2019justifying} to construct the word graph and pre-train CoWPiRec, i.e., ``\textit{Grocery and Gourmet Food}”, ``\textit{Home and Kitchen}”, ``\textit{CDs and Vinyl}”, ``\textit{Kindle Store}” and ``\textit{Movies and TV}”.
  \item \textbf{Downstream datasets}: In the downstream recommendation task, we select another five datasets in the Amazon dataset as cross-domain datasets, namely ``\textit{Industrial and Scientific}”, ``\textit{Prime Pantry}”, ``\textit{Musical Instruments}”, ``\textit{Arts, Crafts and Sewing}”, and ``\textit{Office Products}". We also select a cross-platform dataset, namely \textit{Online Retail}\footnote{https://www.kaggle.com/carrie1/ecommerce-data}, a UK online shopping dataset containing transaction records between 01/12/2010 and 09/12/2011.
\end{itemize}

For all datasets, we remove users and items with fewer than five interactions and arrange the items interacted by users in chronological order following~\cite{hou2022towards}. For item text, we use title, categories, and brand in the Amazon dataset, and item description in the \textit{Online Retail} dataset.

\begin{table}[t]
  \caption{Statistics of the datasets after preprocessing. “Avg. $n$” denotes the average length of item sequences. “Avg. $c$” denotes the average number of tokens in the item text.}
  \label{tab:data}
\tabcolsep=5pt
\centering
\begin{tabular}{lrrrrr}
\toprule
\textbf{Datasets}      & \textbf{\#Users}    & \textbf{\#Items} & \textbf{\#Inters.}  & \textbf{Avg.} n & \textbf{Avg.} c  \cr
\midrule
\textbf{Pre-trained} & 1,361,408  & 446,975 & 14,029,229 & 13.51 & 139.34 \cr
- Food & 115,349 & 39,670 & 1,027,413 & 8.91 & 153.40 \cr
- CDs & 94,010 & 64,439 & 1,118,563 & 12.64 & 80.43 \cr
- Kindle & 138,436 & 98,111 & 2,204,596 & 15.93 & 141.70 \cr
- Movies & 281,700 & 59.203 & 3,226,731 & 11.45 & 97.54 \cr
- Home & 731,913 & 185,552 & 6,451,926 & 8.82 & 168.89 \cr
\midrule
Scientific    & 8,442      & 4,385   & 59,427     & 7.04  & 182.87 \cr
Pantry       & 13,101     & 4,898   & 126,962    & 9.69  & 83.17  \cr
Instruments   & 24,962     & 9,964   & 208,926    & 8.37  & 165.18 \cr
Arts          & 45,486     & 21,019  & 395,150    & 8.69  & 155.57 \cr
Office        & 87,436     & 25,986  & 684,837    & 7.84  & 193.22 \cr
\midrule
Online Retail & 16,520     & 3,469   & 519,906    & 26.90 & 27.80  \cr
\bottomrule
\end{tabular}
\end{table}
\begin{table*}[t!]
\centering
  \caption{Downstream recommendation performance of different models in the fine-tuning setting. The best and the second-best performances are denoted in bold and underlined fonts, respectively. ``H@K” is short for ``HR@K” and ``N@K” is short for ``NDCG@K”, respectively. The subscript `T” denotes that item text is used in the IRL module of the model. The superscripts $^{*}$ and $^{**}$ indicate $p\leq0.05$ and $p\leq0.01$ for the paired t-test of CoWPiRec vs. the best baseline.}
  \label{tab:overall}
\begin{tabular}{clccccccclr}
\toprule
\multicolumn{2}{c}{Setting} & \multicolumn{7}{c}{Baselines} & \multicolumn{2}{c}{Ours} \cr
\cmidrule(lr){1-2} \cmidrule(lr){3-9} \cmidrule(lr){10-11}
 Dataset     & Metric  & SASRec & BERT4Rec & S\textsuperscript{3}Rec\textsubscript{T} & SASRec\textsubscript{T} & ZESRec\textsubscript{T} & UniSRec\textsubscript{T} & MoRec\textsubscript{T} & CoWPiRec\textsubscript{T} & Improv. \cr
\midrule
 \multirow{4}{*}{Scientific}    & H@10   & 0.1063   & 0.0488       & 0.0897 & 0.1163   & 0.1066  & 0.1124  & \underline{0.1174} & \textbf{0.1264}$^{**}$ & +7.67\%  \cr
                          & H@50   & 0.2034   & 0.1185       & 0.1913 & 0.2259  & 0.2095  & 0.2284 & \underline{0.2300} & \textbf{0.2388}$^{**}$  & +3.83\% \cr
                            & N@10 & 0.0552   & 0.0243       & 0.0496 & 0.0631  & 0.0582  & 0.0595 & \underline{0.0635} &  \textbf{0.0664}$^{**}$   & +4.57\% \cr
                             & N@50 & 0.0763   & 0.0393       & 0.0716 & 0.0870  & 0.0808  & 0.0847  & \underline{0.0880} &  \textbf{0.0909}$^{**}$  & +3.30\% \cr
\cmidrule(lr){1-2} \cmidrule(lr){3-9} \cmidrule(lr){10-11}
 \multirow{4}{*}{Pantry}        & H@10   & 0.0493   & 0.0267     & 0.0393 & 0.0603  & 0.0629  & \underline{0.0646} & 0.0639 &  \textbf{0.0679}$^{**}$ &  +5.11\% \cr
                             & H@50   & 0.1333   & 0.0932     & 0.1275 & 0.1676  & 0.1658  & \underline{0.1747} & 0.1682 & \textbf{0.1783}$^{*}$ &  +2.06\%  \cr
                               & N@10 & 0.0219   & 0.0136     & 0.0177 & 0.0295  & 0.0308  & 0.0309 & \underline{0.0310} &  \textbf{0.0320}$^{**}$  &  +3.23\%\cr
                               & N@50 & 0.0399   & 0.0277     & 0.0366 & 0.0528  & 0.0531  & \underline{0.0546} & 0.0535 &  \textbf{0.0559}$^{*}$   & +2.38\% \cr
\cmidrule(lr){1-2} \cmidrule(lr){3-9} \cmidrule(lr){10-11}
\multirow{4}{*}{Instruments}   & H@10   & 0.1126   & 0.0788     & 0.0996 & 0.1175  & 0.109   & 0.1087 & \underline{0.1229} &  \textbf{0.1270}$^{**}$  & +3.34\%  \cr
                             & H@50   & 0.2087   & 0.1485     & 0.1886 & 0.2224  & 0.2044  & 0.2079 & \underline{0.2278} &  \textbf{0.2344}$^{**}$  &  +2.90\%\cr
                               & N@10 & 0.0618   & 0.0579     & 0.0623 & 0.0690  & 0.0649  & 0.0622  &  \underline{0.0717} & \textbf{0.0735}$^{**}$ & +2.51\%  \cr
                               & N@50 & 0.0826   & 0.0728     & 0.0815 & 0.0917  & 0.0855  & 0.0837  & \underline{0.0944} & \textbf{0.0967}$^{**}$  & +2.44\% \cr
\cmidrule(lr){1-2} \cmidrule(lr){3-9} \cmidrule(lr){10-11}
\multirow{4}{*}{Arts}          & H@10   & 0.1074   & 0.0647     & 0.0952 & 0.1078  & 0.1010   & 0.1099  & \underline{0.1101} & \textbf{0.1164}$^{**}$  & +5.72\%  \\
                               & H@50   & 0.1986   & 0.1316     & 0.1815 & 0.2050   & 0.1934  & 0.2118  & \underline{0.2127} & \textbf{0.2231}$^{**}$  & +4.89\% \cr
                              & N@10 & 0.0571   & 0.0403     & 0.0567 & 0.0613   & 0.0568    & 0.0602 & \underline{0.0637} & \textbf{0.0650}$^{**}$   & +2.04\% \cr
                              & N@50 & 0.0769   & 0.0548     & 0.0754 & 0.0825   & 0.0769  & 0.0823  & \underline{0.0860} & \textbf{0.0882}$^{**}$   & +2.56\% \cr
\cmidrule(lr){1-2} \cmidrule(lr){3-9} \cmidrule(lr){10-11}
\multirow{4}{*}{Office}        & H@10   & 0.1064  & 0.0794     & 0.1085 & 0.1043  & 0.0955  & 0.1046  & \underline{0.1096} & \textbf{0.1141}$^{**}$  & +4.11\% \cr
                              & H@50   & 0.1641   & 0.1232      & 0.1683 & 0.1709  & 0.1625  & 0.1751  & \underline{0.1794} & \textbf{0.1867}$^{**}$ &  +4.07\% \cr
                               & N@10 & \textbf{0.0710}   & 0.0573      & 0.0666 & 0.0640   & 0.0567  & 0.0627   & 0.0673 & \underline{0.0703}  & - \cr
                               & N@50 & \underline{0.0835}   & 0.0668      & 0.0797 & 0.0785  & 0.0714  & 0.0780  & 0.0825 & \textbf{0.0861}$^{**}$   & +3.11\% \cr
\cmidrule(lr){1-2} \cmidrule(lr){3-9} \cmidrule(lr){10-11}
 \multirow{4}{*}{Online Retail} & H@10   & 0.1460    & 0.1343      & 0.1433 & 0.1366   & 0.1320   & 0.1444  & \underline{0.1465} & \textbf{0.1515}$^{**}$ & +3.41\%  \cr
                               & H@50   & \underline{0.3872}   & 0.3582      & 0.3762 & 0.3479  & 0.3378  & 0.3653  & 0.3728 & \textbf{0.3928}$^{**}$ & +1.45\%  \cr
                              & N@10 & 0.0671   & 0.0645      & 0.0639 & 0.0666  & 0.0628  & 0.0675 & \underline{0.0712} & \textbf{0.0723}$^{**}$  & +1.54\% \cr
                              & N@50 & 0.1201   & 0.1133      & 0.1146 & 0.1129  & 0.1077  & 0.1158  & \underline{0.1204} & \textbf{0.1247}$^{**}$  & +3.57\%\cr
\bottomrule
\end{tabular}
\end{table*}
\subsubsection{Baselines}
In this paper, we compare CoWPiRec with several baseline methods, including:
\begin{itemize}
    \item \textbf{SASRec}~\cite{kang2018self} uses the self-attention mechanism to aggregate ID-based item representations in the user sequence.
    \item \textbf{BERT4Rec}~\cite{sun2019bert4rec} models user sequence representations based on cloze objective task.
    \item \textbf{SASRec\textsubscript{T}} simply replaces the item ID embedding of SASRec with the item text embedding generated by PLM and maintains the same SRL module.
    \item \textbf{S$^3$Rec}~\cite{zhou2020s3} pre-trains SRL modules with four self-supervised tasks on in-domain data to integrate representations at different levels of features, items, and sequences.
    \item \textbf{ZESRec}~\cite{ding2021zero} obtains item representations using PLM firstly. Then pre-trains the SRL module on data from multiple domains and transfers it to new domains.
    \item \textbf{UniSRec}~\cite{hou2022towards} also obtains item representations using PLM and uses an MoE module to adaptively adjust the representations in different domains. Then the MoE and SRL modules are pre-trained on multi-domain datasets with sequence-item and sequence-sequence contrastive learning tasks.
    \item \textbf{MoRec}~\cite{yuan2023go} performs a joint training on PLM and SRL module with next-item-prediction task. With the item-level supervision signals, the tuned PLM could better adapt to the recommendation task. 
\end{itemize}

Among all the above methods, SASRec and BERT4Rec are ID-based IRL methods. SASRec\textsubscript{T}, ZESRec, UniSRec, MoRec, and our proposed CoWPiRec belong to the text-based IRL methods. 
Different from most baselines, CoWPiRec only pre-trains the IRL module by constructing a word graph containing word-level CF signals and performing a word graph-based pre-training task on datasets from multiple domains. 
Note that we don't compare CoWPiRec with the cross-domain recommendation models since it has been proven that these approaches usually underperform one of our baselines, i.e., UniSRec~\cite{hou2022towards}.

\subsubsection{Evaluation Metric}
We use two widely used evaluation metrics, HR@K and nDCG@K, to evaluate the performance of all models in the next item prediction task on downstream datasets. 
K is set to 10 and 50. 
Following previous work~\cite{kang2018self}, we use the leave-one-out method to construct the dataset. 
Specifically, given a user interaction sequence, the last item is used for testing, the second to last item is used for validation, and the rest of the items are used for training. 
When predicting the next item, we sort all items in the dataset based on the dot-product score. The reported evaluation metrics are the average values of all test users.

\subsubsection{Implementation Details}
We implement CoWPiRec using RecBole~\cite{zhao2021recbole} and transformers~\cite{wolf2020transformers} library. 
For baseline methods, most are implemented by RecBole and we run MoRec with official code\footnote{https://github.com/westlake-repl/IDvs.MoRec}.
During the pre-training stage of CoWPiRec, we construct the word graph by retaining the top 30 co-click words based on their tf-idf scores. 
Item text is tokenized using the BERT tokenizer and we set the maximum length of all item texts to 128. 
Following the BERT masking strategy, we randomly select 15\% of words in the input sequence and replace them with the [MASK] token in 80\% of cases, a random token in 10\% of cases, and leaving them unchanged in 10\% of cases. 
In the word graph modeling step, the number of GNN layers $T$ in the GraphSAGE algorithm is set to 1. 
We use an official checkpoint of BERT in the huggingface hub, i.e., \textit{bert-base-uncased}\footnote{https://huggingface.co/bert-base-uncased} to initialize CoWPiRec's parameters. 
We pre-train CoWPiRec with a batch size of 100 and a learning rate of 5e-5 and use the AdamW optimizer with a linear warm-up rate of 0.1 to update model parameters. 
CoWPiRec is trained for 30 epochs on one Nvidia RTX 3090.

In the fine-tuning setting of the CoWPiRec, we followed~\cite{hou2022towards} and set the number of whitening networks of the MoE module to 8. 
The number of transformer layers and the head of the multi-head self-attention layer in the SRL module are both set to 2. 
For all methods in the downstream recommendation, we use the Adam optimizer and carefully search for hyperparameters, with a batch size of 2048 and early stopping with the patience of 10, using nDCG@10 as the indicator. 
We tune the learning rate in \{0.0003, 0.001, 0.003, 0.01\} and the embedding dimension in \{64, 128, 300\}. 

\subsection{Overall Performance} \label{sec3.2}
\subsubsection{Fine-tuning Setting}
We compare the performance of CoWPiRec with multiple baseline models on five cross-domain datasets and a cross-platform dataset, and the experimental results are shown in Table~\ref{tab:overall}. 

From the results, several observations could be concluded. 
Firstly, Among several baseline methods with ID-based IRL, SASRec achieves better performance when interactions are sufficient while performing poorly on datasets with relatively fewer interactions, e.g., Scientific.
It indicates that the sequential recommender with ID-based IRL heavily relies on ID-based interactions. 
Secondly, The methods with the text-based IRL module effectively improve the performance, especially in datasets that the ID-based model does not specialize in.
Thirdly, with effective joint training on the PLM and the SRL module, MoRec achieves overall better results than other baselines. 
It indicates the significance to enable PLM aware task-specific signals.
While limited by the unsuitable item-level task, the overall performance of MoRec is suboptimal compared to our proposed CoWPiRec.

\begin{table}
\centering
\tabcolsep=2.5pt
\caption{Zero-shot recommendation performance of different models on the downstream datasets. The best and the second-best performances are denoted in bold and underlined fonts, respectively. S\textsuperscript{3}Rec is pre-trained with the same datasets as downstream and other models are pre-trained with Amazon pre-trained data. The superscripts $^{*}$ and $^{**}$ indicate $p\leq0.05$ and $p\leq0.01$ for the paired t-test of CoWPiRec vs. the best baseline.}
\label{tab:zero}
\begin{tabular}{llccccl}
\toprule
Dataset                        & Metric  & ZESRec & S\textsuperscript{3}Rec & UniSRec & MoRec & CoWPiRec \\
\midrule
\multirow{4}{*}{Scientific}    & H@10   & 0.0519 & 0.0025                   & \underline{0.0553} & 0.0481 & \textbf{0.0614}$^{**}$ \\
                               & H@50   & 0.1063 & 0.0158                   & \underline{0.1149} & 0.0943 & \textbf{0.1228}$^{**}$ \\
                               & N@10 & \underline{0.0284} & 0.0011                   & 0.0281 & 0.0222 & \textbf{0.0287}$^{*}$ \\
                               & N@50 & 0.0403 & 0.0039                   & \underline{0.0411} & 0.0324 & \textbf{0.0422}$^{**}$ \\
\midrule
\multirow{4}{*}{Instruments}   & H@10   & \underline{0.0356} & 0.0079                   & 0.0299 & \underline{0.0356} & \textbf{0.0429}$^{**}$ \\
                               & H@50   & 0.0738 & 0.0213                   & \textbf{0.0846} & 0.0649 & \underline{0.0830}  \\
                               & N@10 & \underline{0.0187} & 0.0045                   & 0.0148 & 0.0178 & \textbf{0.0198}$^{**}$ \\
                               & N@50 & \underline{0.0271} & 0.0072                   & 0.0265 & 0.0241 & \textbf{0.0286}$^{**}$ \\
\midrule
\multirow{4}{*}{Online Retail} & H@10   & \underline{0.0375} & 0.0065                   & 0.0369 & 0.0331 & \textbf{0.0440}$^{**}$  \\
                               & H@50   & 0.0780  & 0.0421                   & \underline{0.0814} & 0.0792 & \textbf{0.1011}$^{**}$ \\
                               & N@10 & \underline{0.0180}  & 0.0028                   & 0.0177 & 0.0153 & \textbf{0.0191}$^{**}$ \\
                               & N@50 & 0.0268 & 0.0102                   & \underline{0.0273} & 0.0253 & \textbf{0.0316}$^{**}$ \\
\bottomrule
\end{tabular}
\end{table}

Compared to all baseline models, it is clear that CoWPiRec achieves the best performance in almost all cases.
That demonstrates the effectiveness of incorporating word-level CF signals into the text-based IRL module.
It is worth noting that CoWPiRec trains the MoE module and SRL module from scratch in fine-tuning stage, unlike UniSRec which pre-trains these two modules with mix-domain datasets.
It indicates that the superior result of our model mainly comes from the pre-trained text-based IRL module's ability to capture CF-related information.

\subsubsection{Zero-shot Setting}
For transferable sequential recommenders, the zero-shot performance after transferring to a new domain intuitively reflects the knowledge learned in pre-training. Following the zero-shot recommendation setting in~\cite{ding2021zero}, we directly use the pre-trained checkpoint of transferable sequential recommenders to perform recommendations without any training stage.
Note that in this setting, the model can access all interactions of the user except the last item in the user sequence, but no next-item prediction task training is performed to update the model's parameters.
The experiment results are shown in Table~\ref{tab:zero}. 
From the results, we can conclude several observations. 
Firstly, S\textsuperscript{3}Rec performs poorly in the zero-shot setting.
We speculate the reason is that the modeling procedure of S\textsuperscript{3}Rec's SRL module is different in pre-training and downstream, i.e., bidirectional and unidirectional. 
Secondly, ZESRec, UniSRec, and MoRec perform better than S\textsuperscript{3}Rec, which demonstrates that the pre-training stage contributes to the zero-shot recommendation performance.
Thirdly, CoWPiRec gives clearly better results than other baselines in most cases. 
Note that CoWPiRec is not pre-trained with a recommendation-related task, e.g. next-item-prediction task, It indicates the effectiveness of word graph-based pre-training.
We believe that the significant improvement of CoWPiRec benefits from the word-level CF knowledge learned from the word graph.

\subsection{Cold Start Performance}
\begin{figure}[h]
  \centering
  \includegraphics[width=\linewidth]{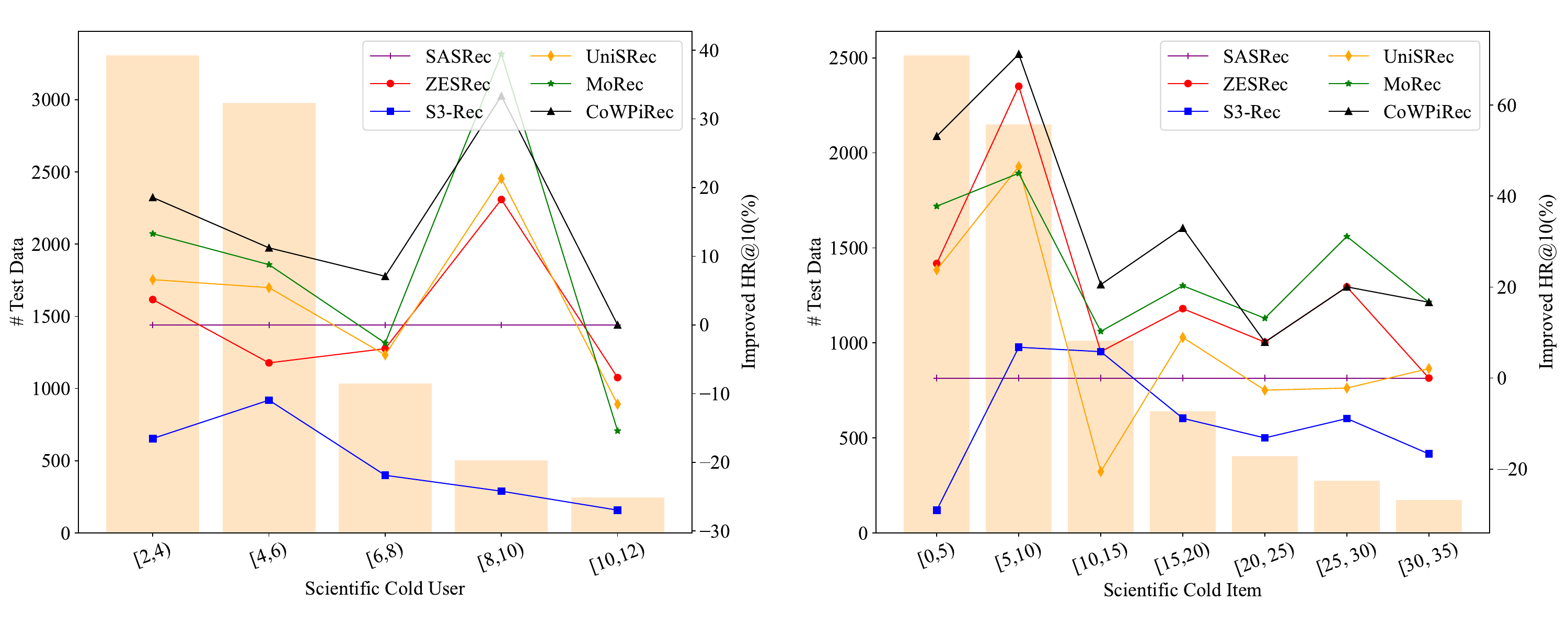}
  \caption{Performance comparison in cold user and cold item experiment on ``Scientific” dataset. The bar graph represents the number of users or items in test data for each group. The line chart represents the improvement ratios for HR@10 compared with SASRec.}
  \label{fig_cold}
\end{figure}
\begin{figure*}[t]
  \centering
  \includegraphics[width=0.95\linewidth]{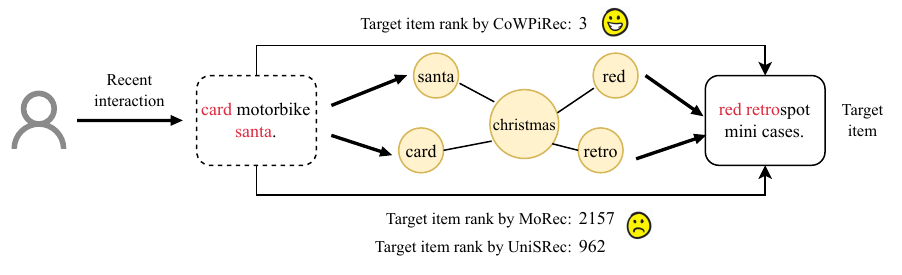}
  \caption{The interaction history of a user in the ``Online Retail” dataset, a sub-graph of our constructed word graph, and the rank results of the target item of models in the zero-shot setting. The word ``card” and ``santa” have co-click relationships with ``retro” and ``red” in the word graph. CoWPiRec utilizes the word-level CF signal learned from the word graph and captures ``red” and ``retro” in the target item. Therefore, CoWPiRec ranks the target item at a high position and achieves a clearly better performance than other models.}
  \label{fig_case}
\end{figure*}
One goal of the transferable sequential recommender is to alleviate the cold start issue in new domains. 
We evaluate CoWPiRec's performance compared to baseline models on the cold start setting from two perspectives: cold users and cold items. 
Specifically, for cold user experiments, we group the users in the test set based on the number of their interactions in the training set.
For cold item experiments, we split the test set based on the target item's popularity in the training set. 
We present the relative improvement of CoWPiRec and several baselines over SASRec in terms of HR@10, as shown in Figure~\ref{fig_cold}.

From the result, several observations can be concluded.
Firstly, CoWPiRec achieves the most improvement over SASRec in most user groups while
other baseline models underperform SASRec in some groups.
Secondly, in the cold item experiment, CoWPiRec significantly improves the performance in most item groups, especially in the items group that are less interacted with by users, i.e., group [0,5) and [5,10).
The experiment result demonstrates that CoWPiRec can effectively alleviate the cold-start issue in cross-scenario recommendations utilizing the item representations capturing the word-level CF signals.

\subsection{Case Study}
From the experimental results in section~\ref{sec3.2}, we can see that CoWPiRec achieves significantly better performance than other methods in most cases.
Since we did not perform cross-domain pre-training for the SRL module, or even don't leverage it (i.e., zero-shot setting).
We believe that the performance improvement of CoWPiRec mainly comes from the ability learned in the pre-training stage to capture word-level CF signals.
We will show a case to illustrate how CoWPiRec leverages the knowledge learned from the word graph-based pre-training to improve the performance of downstream recommendation tasks.

In the case shown in Figure~\ref{fig_case}, CoWPiRec ranks the ground-truth next item at the 3rd position without any in-domain user interaction data training (i.e., zero-shot setting).
It is significantly better than two strong baselines, i.e., MoRec and UniSRec.
We believe CoWPiRec achieves significantly better ranking performance by capturing the word-level user preferences, i.e., the words ``santa” and ``card” in the recent interaction and the words ``red” and ``retro” in the target item.
We can find co-click relationships with similar word-level preferences in the word graph. 
It indicates that CoWPiRec learns these word-level CF signals from word graph-based pre-training and applies the learned knowledge to the recommendation task in downstream datasets. 

\section{Conclusion}
In this paper, we proposed a transferable item representation learning framework, named CoWPiRec. 
Different from previous transferable sequential recommenders that typically utilize the text-based IRL module as an offline feature extractor and learn a universal SRL module,
we focus on incorporating recommendation knowledge into the text-based IRL module allowing it to capture CF signals.
Considering the item-level CF signal is not suitable for the widely used text-based IRL module, i.e., PLM.
We first construct a word graph fused with CF signals by collecting co-click word pairs and then integrating these signals into the PLM via a word-level pre-training task.
With the ability to capture word-level recommendation information, CoWPiRec can even perform recommendations with a simple SRL module without trainable parameters, i.e., mean pooling.
Furthermore, combining CoWPiRec with the SRL module and performing downstream training can achieve significantly better performance compared with state-of-the-art transferable sequential recommenders. 
Note that the SRL module used in the experiment is not tailored for CoWPiRec and just follows a previous architecture.
It leaves us a future work of exploring a sophisticated SRL to improve the performance of CoWPiRec.

\bibliographystyle{IEEEtran}
\bibliography{IEEEabrv,sample}

\end{document}